\documentclass[12pt]{article}
\usepackage{latexsym,graphicx,url}

\title{Efficient Tree Layout in a Multilevel Memory Hierarchy%
  \footnote{A preliminary version of this paper appeared in
    ESA 2002~\protect\cite{Bender-Demaine-Farach-Colton-2002-tree-layout}.}}

\author{%
  Stephen Alstrup%
    \thanks{IT University of Copenhagen, Glentevej 65-67,
      DK-2400 Copenhagen NV, Denmark;
      email: \{\texttt{stephen}, \texttt{theis}\}\protect\url{@itu.dk}.}
\and
  Michael A. Bender%
    \thanks{Department of Computer Science,
      State University of New York, Stony Brook, NY 11794-4400, USA;
      email: \protect\url{bender@cs.sunysb.edu}.
      Supported in part by Sandia National Laboratories and
      the National Science Foundation grants EIA-0112849 and CCR-0208670.}
\and
  Erik D. Demaine%
    \thanks{MIT Laboratory for Computer Science,
      200 Technology Square, Cambridge, MA 02139, USA;
      email: \protect\url{edemaine@mit.edu}.
      Supported in part by NSF Grant EIA-0112849.}
\and
  Martin Farach-Colton%
    \thanks{Department of Computer Science,
      Rutgers University, Piscataway, NJ 08855, USA;
      email: \protect\url{farach@cs.rutgers.edu}.
      Supported in part by NSF Grant CCR-9820879.}
\and
  Theis Rauhe%
    \footnotemark[2]
\and
  Mikkel Thorup%
    \thanks{AT\&T Labs---Research, Shannon Laboratory, Florham Park, NJ 07932,
      USA; email: \protect\url{mthorup@research.att.com}.}
}

\date{}

\newif\ifDP
\DPtrue

\advance \topmargin by -\headheight
\advance \topmargin by -\headsep
\evensidemargin \oddsidemargin
\let\latexcite=\cite
\let\latexref=\ref
\def\ref{\nolinebreak\latexref}

{\makeatletter
 \gdef\xxxmark{%
   \expandafter\ifx\csname @mpargs\endcsname\relax 
     \expandafter\ifx\csname @captype\endcsname\relax 
       \marginpar{xxx}
     \else
       xxx 
     \fi
   \else
     xxx 
   \fi}
 \gdef\xxx{\@ifnextchar[\xxx@lab\xxx@nolab}
 \long\gdef\xxx@lab[#1]#2{{\bf [\xxxmark #2 ---{\sc #1}]}}
 \long\gdef\xxx@nolab#1{{\bf [\xxxmark #1]}}
}

{\makeatletter \gdef\fps@figure{!htbp}}

\newtheorem{theorem}{Theorem}
\newtheorem{lemma}{Lemma}
\newtheorem{claim}{Claim}
\newtheorem{corollary}[claim]{Corollary}

\makeatletter
\def\GrabProofArgument[#1]{ #1: \egroup\ignorespaces}
\def\proof{\noindent\textbf\bgroup Proof%
           \@ifnextchar[{\GrabProofArgument}{: \egroup\ignorespaces}}

\makeatother

\def\hyperspc{\kern -0.22em}

\newcommand{\lhceil}{\lceil \hyperspc \lceil}
\newcommand{\rhceil}{\rceil \hyperspc \rceil}

\let\log=\lg

\begin{document}
\maketitle

\begin{abstract}
  We consider the problem of laying out a tree with fixed parent/child
  structure in hierarchical memory.  The goal is to minimize the
  expected number of block transfers performed during a search along a
  root-to-leaf path, subject to a given probability distribution on
  the leaves.  This problem was previously considered by Gil and Itai,
  who developed optimal but slow algorithms when the
  block-transfer size $B$ is known.
  \ifDP
  We present faster but approximate
  algorithms for the same problem; the fastest such algorithm runs in
  linear time and produces a solution that is within an additive
  constant of optimal.  

  In addition, we
  \else
  We
  \fi
  show how to extend any approximately optimal
  algorithm to the \emph{cache-oblivious} setting in which the
  block-transfer size is unknown to the algorithm.  The query
  performance of the cache-oblivious layout is within a constant
  factor of the query performance of the optimal known-block-size
  layout.  Computing the cache-oblivious layout requires only
  logarithmically many calls to the layout algorithm
  for known block size\ifDP; in
  particular, the cache-oblivious layout can be computed in $O(N \log
  N)$ time, where $N$ is the number of nodes\fi.
  \ifDP

  \fi
  Finally, we analyze two greedy strategies, and show that they have
  a performance ratio between $\Omega(\log B / \log \log B)$ and $O(\log B)$
  when compared to the optimal layout.
\end{abstract}

\section{Introduction}

The B-tree~\cite{Bayer-McCreight-1972} is the classic optimal search
tree for external memory, but it is only optimal when accesses are
uniformly distributed.  In practice most distributions are nonuniform,
e.g., distributions with heavy tails arise almost universally
throughout computer science.

Consequently, there is a large body of work on optimizing search trees
for nonuniform distributions in a variety of contexts:
  \begin{enumerate}
  \item \emph{Known distribution on a RAM} --- optimal binary search 
        trees~\cite{Aho-Hopcroft-Ullman-1974, Knuth-1968-volume-3}
        and variations~\cite{Hu-Tucker-1971-SIAM}, and Huffman codes
        \cite{Huffman-1952}.
  \item \emph{Unknown distribution on a RAM} ---
        splay trees and variations~\cite{Iacono-2001-splay, Sleator-Tarjan-1985-splay}.
  \item \emph{Known distribution in external memory} ---
        optimal binary search trees in the 
        HMM model~\cite{Thite-2001}.
  \item \emph{Unknown distribution in external memory} ---
        alternatives to splay trees~\cite{Iacono-2001-splay}.%
    \footnote{Although~\cite{Iacono-2001-splay}
      does not explicitly state its results
      in the external-memory model, its approach easily applies to this
      scenario.}
  \end{enumerate}

\paragraph{Fixed Tree Topology.}
Search trees frequently encode decision trees that cannot be
rebalanced because the operations lack associativity.  Such trees naturally
arise in the context of string or geometric data, where each node represents a
character in the string or a geometric predicate on the data.
Examples of such structures include tries, suffix trees,
Cartesian trees, k-d trees and other BSP trees, quadtrees, etc.
Almost always their contents are not uniformly distributed, and
often these search trees are unbalanced.

The first question is how to optimize these fixed-topology trees when
the access distribution is known.  On a RAM there is nothing to
optimize because there is nothing to vary.  In external memory,
however, we can choose the layout of the tree structure in memory,
that is, which nodes of the tree are stored in which blocks in memory.
This problem was first considered by Gil and
Itai~\cite{Gil-Itai-1999}.  Among other results described below, they
presented a dynamic-programming algorithm for optimizing the partition
of the $N$ nodes into blocks of size $B$, given the probability
distribution on the leaves.  The algorithm runs in $O(N B^2)$ time
and uses $O(B \log N)$ space.

This problem brings up important issues in external-memory algorithms
because when trees are unbalanced or distributions are skewed, there
is even more advantage to a good layout.  Whereas uniform
distributions lead to B-trees, which save a factor of only $\lg B$
over standard $(\lg N)$-time balanced binary trees, the savings grow
with nonuniformity in the tree.  In the extreme case of a
linear-height tree or a very skewed distribution, an optimal memory layout
obtains a factor of $B$ savings over a na\"{\i}ve layout.

\paragraph{Cache-Oblivious Algorithms.}
Recently, there has been a surge of interest in data structures for
multilevel memory hierarchies.  Frigo, Leiserson, Prokop, and
Ramachandran~\cite{Frigo-Leiserson-Prokop-Ramachandran-1999,Prokop-1999}
introduced the notion of \emph{cache-oblivious algorithms}, where the objective
is to obtain asymptotically optimal memory performance for all possible values
of the memory-hierarchy parameters (block size and memory-level size).
As a consequence, such algorithms tune automatically to arbitrary
memory hierarchies with an arbitrarily many memory levels.  Examples
of cache-oblivious \emph{data structures} include cache-oblivious
B-trees~\cite{Bender-Demaine-Farach-Colton-2000} and its
simplifications~\cite{Bender-Duan-Iacono-Wu-2002,
Brodal-Fagerberg-Jacob-2002, Rahman-Cole-Raman-2001}, cache-oblivious
persistent trees~\cite{Bender-Cole-Raman-2002}, cache-oblivious
priority queues~\cite{Arge-Bender-Demaine-Holland-Minkley-Munro-2002},
and cache-oblivious linked lists~\cite{Bender-Cole-Demaine-Farach-Colton-2002-scans}.
However, all of these data structures assume a uniform distribution on
operations.

\paragraph{Our Results.}
We obtain several results about tree layout:

\begin{enumerate}
\item We present fast dynamic-programming algorithms for tree layout
  with known block size that trade off between layout quality and running time.
  The fastest algorithm runs in $O(N)$ time and computes a layout that is
  within an additive $1+\delta$ of optimal for any $\delta > 0$.
\item We develop a general technique called Split-and-Refine for converting any
  family of layouts with known block size into a layout with unknown block
  size, while increasing the expected block cost by at most a constant factor.
  Using the previous result, we obtain a running time of $O(N \lg N)$.
\item In addition, we show how to adapt this technique to other objective
  functions, specifically, minimizing the maximum block cost.
\item We analyze two natural greedy algorithms for tree layout with known
  block size.  We show that their performance can be as bad as a factor of
  $\Omega(\log B / \log \log B)$ away from optimal, but is no more than
  $O(\log B)$ away from optimal.
\end{enumerate}

In the conference version of this
paper~\cite{Bender-Demaine-Farach-Colton-2002-tree-layout},
two incorrect claims were made: that
a simple greedy algorithm is within an additive $1$ of optimal for known
block size, and that a particular recursive greedy partition of the
tree yields a cache-oblivious algorithm within a multiplicative $O(1)$
of optimal.  This paper gives corrected versions of these claims.

\paragraph{Related Work.}

In addition to the result mentioned above, Gil and
Itai~\cite{Gil-Itai-1995,Gil-Itai-1999} consider other algorithmic
questions on tree layouts.  They prove that minimizing the number of
distinct blocks visited by each query is equivalent to minimizing the
number of block transfers over several queries; in other words,
caching blocks over multiple queries does not change the optimal
solutions.  Gil and Itai also consider the situation in which the
total number of blocks must be minimized (the external store is
expensive) and prove that optimizing the tree layout subject to this
constraint is NP-hard.  In contrast, with the same constraint, it is
possible to optimize the expected query cost within an additive $1/2$
in $O(N \log N)$ time and $O(B)$ space.  This algorithm is a variant
of their polynomial-time dynamic program for the unconstrained
problem.

Clark and Munro \cite{Clark-Munro-1996, Clark-1996}
consider a worst-case version
of the problem in which the goal is to minimize the maximum number of
block transfers over all queries, instead of minimizing the expected number of
block transfers.  They show how the exact optimal layout can be computed
in $O(N)$ time for a known block size~$B$.
We show how to extend this result to the cache-oblivious setting.

\section{Basics}

We begin with a few definitions and structural properties.

In the \emph{tree layout} (or \emph{trie layout}) problem,
we are given a fixed-topology tree with a known probability distribution on
the leaves.  When the layout algorithm knows the memory block size $B$,
the goal is to produce a \emph{layout}, which clusters nodes into memory
blocks.
In the cache-oblivious model, where the layout algorithm does not know~$B$,
the goal is to determine a \emph{cache-oblivious layout}, which specifies
an order of the nodes in memory.
A cache-oblivious layout must be efficient no matter how the
ordering of nodes is partitioned into consecutive blocks of size $B$,
and for all values of~$B$.

The \emph{expected block cost} of a layout is the expected number of blocks
along the root-to-leaf path for a randomly selected leaf.
The \emph{optimal layout with block size $B$} minimizes the expected block cost
over all possible layouts with block size~$B$.

A simple but useful idea is to propagate the probability distribution on
leaves up to internal nodes.
Define the \emph{probability of an internal node} to be the
probability that the node is on a root-to-leaf path for a randomly
chosen leaf; that is, the probability of an internal node is the sum
of the probabilities of the leaves in its subtree.  These
probabilities can be computed and stored at the nodes in linear time.


All of our algorithms are based on the following structural lemmas:

\begin{lemma}[Convexity Lemma \cite{Gil-Itai-1999}] \label{convexity}
  Any fixed-topology tree has an optimal layout that is \emph{convex}, i.e.,
  in which every block forms a connected subtree of the original tree.
\end{lemma}

\begin{lemma}[Monotonicity Lemma] \label{monotonicity}
  The expected search cost for the optimal tree layout with block size $B-1$
  is no greater than for the optimal tree layout with block size $B$.
\end{lemma}

\begin{proof}
  The blocks of an optimal layout with block size $B$ need not be full.
  In particular, each node could store just $B-1$ elements.
\end{proof}

\begin{lemma}[Smoothness Lemma] \label{smoothness}
  An optimal tree layout with block size $B/2$
  has an expected search cost of no more than twice
  that of an optimal layout with block size~$B$.
\end{lemma}

\begin{proof}
  Consider an optimal convex layout with block size~$B$.
  Partition each block arbitrarily into two pieces of size $B/2$.
  This new layout has at most twice as many block transfers as the optimal
  layout with block size~$B$, and so the optimal layout with block size $B/2$
  does at least this well.
\end{proof}

\section{Performance of Greedy}

In this section we analyze the performance of two greedy heuristics
for tree layout with known block size~$B$.
Despite the natural appeal of these algorithms,
we show that their performance can be roughly a $\log B$ factor from optimal.

The most natural greedy heuristic, \emph{Weight-Greedy}, incrementally grows a
\emph{root block}, that is, the block containing the root of the tree.
Initially, the root block contains just the root node; then the heuristic
repeatedly adds the maximum-probability node not already in the root block
(which is necessarily adjacent to the subtree so far).
When the root block fills, i.e., contains $B$ nodes,
the heuristic conceptually removes its nodes from the tree
and lays out the remaining subtrees recursively,
storing each subtree in a separate memory region.

The \emph{DFS-Greedy} heuristic orders the nodes according to a locally
greedy depth-first search, and then partitions the nodes into consecutive
blocks of size~$B$.
More precisely, the nodes are ordered according to when they are
traversed by a depth-first search that visits the children
of a node by decreasing probability.
This layout is not necessarily convex.

We give a near-tight analysis of the worst-case ratio between the expected
block cost in either greedy algorithm and the expected block cost in the
optimal layout.  Specifically, we prove that this ratio is between
$\Omega(\log B / \log \log B)$ and $O(\log B)$.

\subsection{Lower Bound of \boldmath$\Omega(\log B / \log \log B)$}

\begin{theorem}
  \label{thm:logoverlogloglowerbound}
  The expected block cost of either greedy layout (Weight-Greedy
  or DFS-Greedy) can be a factor of $\Omega(\log B / \log \log B)$
  more than the expected block cost of the optimal layout.
\end{theorem}

\begin{proof}
We exhibit at tree $ T $
with $N \geq B ^ 2 $
nodes, for which the expected greedy block cost is 
$\Theta (\log B/\log\log B) $
and the optimal block cost is
$\Theta (1) $.
Then we show how to replicate this tree 
so that the expected greedy block cost is
$\Theta (\log N/ \log B) $
and the optimal block cost is
$\Theta (\log N / \log\log B) $.

We build tree $T$
by starting with a complete $B$-node tree $T'$
with fanout  $\Theta (\log B) $ and height $\Theta (\log B/\log\log B) $,
in which all nodes at each level have equal probability.
We next augment tree $T' $ by attaching an \emph{escape path} of length~$B$
to every node.
The probability of the first node on an escape path is slightly 
higher than the probability of its sibling nodes.
(For example, we can let the $\log B$ children in $T'$ of a parent node
have probability $1/(2+\log B)$ times the probability of the parent,
and then we let the escape path have probability $2/(2+\log B)$ times the
probability of the parent.)
Thus, greedy favors all nodes along the escape paths instead of any of the
adjacent children.
This construction of $T$ has $\Theta(B^2)$ nodes.

The optimal layout assigns the original tree $T'$
to its own block, and
assigns each escape path to its own block.
Because there are $\Theta(B^2)$ nodes,
there are $\Theta(B)$ blocks.
The expected search cost is~$2$.

Greedy performs worse.
Because each escape path has length $B$,
and because greedy favors the escape paths,
each node in the tree $T'$
is in its own block.
Thus, a search pays a cost of $1$ for each level in $T'$
visited before the search chooses an escape path and leaves~$T'$.
The probability that a random search reaches the leaves in $T'$ is
$(1 - 1/\log B)^{\Theta (\log B/\log\log B)} \approx (1/e)^{1/\log \log B}$,
which is at least $1/e$ (and in fact much closer to~$1$).
Thus, at least a constant fraction of searches reach the bottom of the tree,
visiting one block for each of $\Theta(\log B/\log \log B)$ levels.

In summary,
for a tree $T$ with $\Theta(B^2)$ nodes,
optimal has expected block cost of~$2$,
whereas greedy has expected block cost
$\Theta (\log B/\log\log B)$.

We can grow tree $T$ to have arbitrary many nodes
by splitting each leaf into many leaves all with same parent.
This change can only affect the costs of greedy and optimal by at most~$1$.
Thus, the ratios between the costs remains $\Theta(\log B / \log \log B)$,
but the costs do not grow with~$N$.

Alternatively, we can replicate the tree $T$ so that the block costs
grow with~$N$.
Specifically, we attach another copy of $T$ to the end of each escape path,
and we repeat this process any desired number of times.
Each iteration increases the size of $T$ by roughly a factor of~$B$.
The result is that the optimal expected search cost is
$S(N) = S(N/B) + \Theta(1) = \Theta(\log N / \log B)$,
whereas the greedy expected search cost is
$S(N) = S(N/B) + \Theta(\log B/\log\log B) = \Theta(\log N/ \log\log B)$.
\end{proof}

\subsection{Upper Bound of \boldmath$O(\lg B)$}

We now prove that no tree is much worse than the class of examples in the
previous section:

\begin{theorem}
  \label{thm:logupperbound}
  The expected block cost of either greedy layout (Weight-Greedy or DFS-Greedy)
  is within a factor of $O(\lg B)$ of optimal.
\end{theorem}

To simplify boundary cases, we preprocess the tree as follows:
to each to leaf, we attach $2 B$ children called \emph{hair nodes}.
The probability of each hair node is $1/(2B)$ of the probability of the leaf.
This preprocessing only increases the expected search cost in greedy,
and it increases the optimal search cost by at most~$1$ (because we could put
each hair node in its own block, and the optimal solution can only be better).

We partition the tree into \emph{chunks} as follows,
and treat each chunk separately.
We grow the first chunk starting at the root.
The \emph{probability of a chunk} is the probability of entering its root;
for the first chunk, this probability is~$1$.
Define the \emph{relative probability} of a node in a chunk to be
the probability of entering that node divided by the probability of the chunk.
We grow a chunk by repeating the following process:
for every node in the chunk that has relative probability more than $1/2B$,
we add the children of that node to the chunk.
When this process terminates, we have completed a chunk; we conceptually remove
those nodes and recurse on the remainder of the tree to complete the
partition into chunks.

As a postprocessing step, we demote from chunkhood any chunk that consists
solely of a hair node, leaving that hair node separate from all chunks.
The reason for the careful handing of hair nodes is to ensure that
all of the leaves of a chunk have relative probability at most $1/2B$.
Furthermore, the parent of each leaf of a chunk, which we call a \emph{twig
node}, has relative probability more than $1/2B$.

We prove two lower bounds on the optimal block partition, and one
upper bound on the greedy block partition.

\begin{claim}[Lower Bound 1] \label{LB1}
  Consider the optimal block partition of the tree.
  The number of blocks along a path from the root of the tree
  to a leaf of the tree is at least the length of that path divided by~$B$.
\end{claim}

\begin{proof}
  We need to visit every node along the path from the root of the tree
  to the leaf of the tree.  If there are $P$ nodes along the path,
  the best we could hope for is that every block contains $B$ of these
  $P$ nodes along the path.
  Thus, the number of blocks visited is at least $\lceil P/B \rceil$.
  This lower bound can only be larger than $P/B$, which is the claimed bound.
\end{proof}

The next lower bound first considers each chunk separately,
and then combines the chunk-by-chunk bounds to apply to the entire tree.
If we counted the number of blocks along a root-to-leaf path separately for
each chunk, and then added these counts together, we might double-count blocks
because a single block can overlap multiple chunks.
Instead, we count the number of \emph{block transitions} along a path,
i.e., the number of traversed edges that straddle two blocks, which is exactly
$1$ smaller than the number of blocks along that path.
Now we can count the number of block transitions separately for the subpath
within each chunk, then add these counts together, and the resulting total
only underestimates the true number of block transitions.

\begin{claim}[Lower Bound 2 Within Chunk] \label{LB2C}
  Consider the optimal block partition of the tree.
  Within any chunk, the expected number of block transitions
  from the root of the chunk to a leaf of the chunk is at least~$1/2$.
\end{claim}

\begin{proof}
  The memory block containing the root of the chunk has at most $B$ nodes,
  so it can contain at most $B$ leaves of the chunk.
  Each leaf of the chunk has relative probability at most $1/2B$,
  so $B$ leaves of the chunk have total relative probability at most~$1/2$.
  Thus, the leaves of the chunk that do not fit in the root block
  of the chunk have a total relative probability of at least~$1/2$.
  For these leaves, the number of block transitions within the chunk
  is at least~$1$.
  Therefore, the expected number of block transitions within the chunk
  is at least~$1/2$.
\end{proof}

Now we combine the estimates at each chunk to form a bound on the entire tree.
Define the \emph{expected chunk count} to be the expected number of chunks
along a root-to-leaf path.

\begin{corollary}[Lower Bound 2] \label{LB2}
  Consider the optimal block partition of the tree.
  The expected number of blocks along a path from the root of the tree
  to a leaf of the tree is at least half the expected chunk count.
\end{corollary}

\begin{proof}
  Label the chunks from $1$ to $k$.
  Consider a random root-to-leaf path.
  Define random variable $X_i$ to be the number of block transitions along the
  path that occur within chunk $i$, i.e., the number of block transitions
  along the subpath entirely contained in chunk $i$.
  If the path does not visit chunk $i$, then $X_i$ is~$0$.
  Conditioned upon the path visiting chunk $i$, $X_i$ is at least~$1/2$,
  by Claim \ref{LB2C}.

  Define random variable $X$ to be $\sum_{i=1}^k X_i$,
  which counts all block transitions strictly within chunks, but ignores
  block transitions that align with chunk boundaries.
  Thus, $X$ is a lower bound on the number of block transitions along the path.
  By linearity of expectation, $E[X]$ equals $\sum_{i=1}^k E[X_i]$.
  By the argument above, $E[X_i]$ is at least $1/2$ times the probability of
  entering chunk~$i$.  Thus, $E[X]$ is at least half the expected chunk count.

  This lower bound ignores any additional cost potentially caused by hair nodes
  that do not belong to chunks, which would only further increase
  the lower bound.
\end{proof}

Now we establish an upper bound on either greedy block partition.

\begin{claim}[Upper Bound Within Chunk] \label{UBC}
  Consider either greedy block partition of the tree
  (Weight-Greedy or DFS-Greedy).
  Within any chunk, the number of blocks along a path
  from the root of the chunk to a leaf of the chunk
  is at most the length of that path divided by $B$, plus $2 \lg B + 7$.
\end{claim}

\begin{proof}
  We divide the chunk into \emph{strata} based on the nearest power of two
  of the relative probability of a node.
  More precisely, define \emph{stratum $i$} of the chunk
  to contain the nodes with relative probability at most $1/2^i$
  and more than $1/2^{i+1}$.
  We consider strata $0$, $1$, $2$, \dots, $\lceil \lg 2B \rceil$.
  Thus, there are $1 + \lceil \lg 2B \rceil \leq \lg B + 3$ strata.
  Leaf nodes may have sufficiently small probability to be excluded from
  all strata.  However, the strata partition all nonleaf nodes of the chunk
  (i.e., down to the twig nodes).

  We claim that each stratum is a vertex-disjoint union of paths.
  Consider any node in stratum~$i$, which by definition has relative
  probability at most $1/2^i$.
  At most one child of that node can have relative probability more than
  $1/2^{i+1}$.  Thus, at most one child of each node in stratum~$i$
  is also in stratum~$i$, so each connected component of stratum~$i$
  is a path.  We call these connected components \emph{subpaths}.

  Any path from the root of the chunk to a leaf of the chunk
  starts in stratum $0$, and visits some of the strata in increasing order.
  The path never revisits a stratum that it already left, because
  the probabilities of the nodes along the path are monotonically decreasing.
  Thus, the path passes through at most $\lg B + 3$ strata.

  When the greedy heuristic adds a node from a particular stratum
  to a block, it continues adding all nodes from that stratum to the
  block, until either the block fills or the subpath within the stratum
  is exhausted.
  Thus, after an initial startup of at most $B$ nodes in a subpath,
  greedy clusters the remainder of the subpath into full blocks of size~$B$.
  Because of potential roundoff at the top and the bottom of this subpath,
  the number of blocks along this subpath is at most the length of the
  subpath divided by $B$, plus $2$.
  In addition, the leaves of the chunk may not belong to any stratum,
  so we may visit one additional block after visiting the strata.
  Summing over all strata, the total number of blocks along the path from
  the root of the chunk to a leaf of the chunk is at most the length of the
  path divided by $B$, plus $2 \lg B + 6$ ($2$ per stratum),
  plus $1$ (for a leaf).
\end{proof}

\begin{corollary}[Upper Bound] \label{UB}
  Consider either greedy block partition of the tree
  (Weight-Greedy or DFS-Greedy).
  The number of blocks along a path from the root of the tree
  to a leaf of the tree is at most the length of that path divided by $B$,
  plus $2 \lg B + 8$ times the expected chunk count.
\end{corollary}

\begin{proof}
  We follow the same proof outline as Corollary \ref{LB2}
  except for three difference.
  First, we plug in Claim \ref{UBC} instead of Claim \ref{LB2C}.
  Second, before we counted the number of block transitions
  along the path, to avoid over-counting for a lower bound, whereas here we
  count the number of blocks along a path,
  to avoid under-counting for an upper bound.
  Third, we add an additional $1$ to the bound
  because of potential hair nodes separate from all chunks.
\end{proof}

Finally, we combine the lower bounds in Lemma \ref{LB1} and Corollary
\ref{LB2}, and the upper bound in Corollary \ref{UB},
to prove Theorem \ref{thm:logupperbound}:

\medskip

\begin{proof}[of Theorem \ref{thm:logupperbound}]
By Corollary \ref{UB},
the expected number of blocks along a root-to-leaf path is at most
the expected path length divided by~$B$,
plus $2 \lg B + 8$ times the expected chunk count.
By Lemma \ref{LB1}, this expected cost is at most the
optimal expected search cost,
plus $2 \lg B + 8$ times the expected chunk count.
Furthermore, by Corollary \ref{LB2}, the optimal expected search cost
is at least half the expected chunk count.
Therefore, the ratio of the greedy expected search cost over the
optimal expected search cost is at most
$1 + 2 (2 \lg B + 8)$.
That is, greedy performs within a factor of $4 \lg B + 17$ from optimal.
\end{proof}

\ifDP
\section{Faster Algorithms for Known Block Size}
\label{DP}

In this section, we show how to approximate the minimum
expected block cost in a layout with known block size~$B$.
All of our algorithms are dynamic programs and build off of the original
dynamic program in \cite{Gil-Itai-1999}, which runs in $O(N B^2)$ time.%
\footnote{The analysis in \cite{Gil-Itai-1999} is slightly loose,
  claiming an $O(N B^2 \log \Delta)$ bound
  where $\Delta$ is the maximum degree of a node,
  but an $O(N B^2)$ bound holds of the same algorithm.}
We begin with a brief description of this dynamic program in
Section \ref{Gil and Itai}.
As we loosen the constraints of the layout and allow further approximation,
we obtain faster running times.  In Section \ref{B reduction} we reduce the
effective problem size and running time by a factor of $B$ at the cost of an
additive~$1$ in memory transfers.  Unfortunately, this reduction cannot be
applied a second time.  Finally in Section \ref{fancy DP} we reduce the running
time to $O(N)$ by approximating the solution to the dynamic program,
incurring an additional $\delta$ additive overhead for any $\delta > 0$.

\subsection{\boldmath Gil and Itai's Algorithm in $O(N B^2)$ Time}
\label{Gil and Itai}

We start with a review of Gil and Itai's dynamic program because it serves as
our starting point for faster algorithms.

The basic idea in all of the dynamic programs comes from the Convexity Lemma
(Lemma~\ref{convexity}): try all possible root blocks, recursively compute
layouts of the remaining subtrees, compute the total expected cost, and find
the best overall layout.
The simplest dynamic program based on this idea has a subproblem for each
subtree rooted at a node~$v$.  Thus, there are $N$ subproblems.
Unfortunately, the running time of this algorithm has an exponential dependence
on~$B$ because there are many choices for the root block.
For example, when the root node has $\Theta(N)$ children, we must select $B-1$
out of these children for inclusion in the root block, which requires
${\Theta(N) \choose \Theta(B)} = N^{\Theta(\min\{B,N-B\})}$ time.

To reduce the dependence on $B$, we strengthen the induction hypothesis
by adding more subproblems.  Namely, for each node $v$ and for each \emph{root
capacity} $0 \leq K \leq B$, we consider laying out the subtree rooted at $v$
with the additional constraint that the root block stores at most
$K$ nodes.
A root capacity of $K < B$ represents that a portion of a block is reserved for
the root block of this subtree, but the rest of the block is used by parent or
sibling subproblems.
As before, all blocks except the root block have capacity for
up to $B$ nodes.
Now there are $N (B+1)$ subproblems.

The \emph{expected block cost of a subproblem} rooted at node $v$ is the
expected number of blocks within the subtree rooted at $v$, excluding the
root block, that are visited by a random root-to-leaf path in the entire tree.
Notice that many such root-to-leaf paths do not visit any blocks within the
subtree rooted at $v$; thus, the expected block cost of a subproblem rooted at
$v$ is effectively weighted by the probability of node~$v$.
Hence, the expected block cost of a subproblem measures the contribution of the
subproblem to the expected block cost of the entire layout.
The cost of accessing the root block is not counted in the cost of a
subproblem, because this cost has already been paid by the parent.
Thus, the expected block cost of a subproblem rooted at a leaf is~$0$.
At the other extreme, the expected block cost of the subproblem at the root of
the tree and with root capacity $0$ is the expected block cost of the overall
layout.

To solve the subproblem rooted at node $v$ and with root capacity $K = 0$,
we recursively solve the subproblem rooted at node $v$ with root capacity
$K' = B$.  This recursion corresponds to terminating one root block and
starting a new root block of full capacity.
The expected block cost of the original subproblem is the expected block cost
of the recursive subproblem plus the probability of node $v$, i.e.,
the expected cost of accessing the new root block.
This sum is the only place in the dynamic program
where the overall cost is increased.

To solve the subproblem rooted at node $v$ and with root capacity $K \geq 1$,
we first place the root node in the root block, consuming one unit
of the root capacity.  Then we guess how much of the remaining root capacity
$K-1$ should be reserved for each of the children of the root node.
In other words, we consider all partitions
$K-1 = K_1 + K_2 + \cdots + K_d$, where each $K_i \geq 0$ and $d$ is the (out)
degree of the root node.
Then we recursively compute the optimal layout for each of the $d$ children
subtrees, where child $i$ is assigned root capacity $K_i$ representing
the reserved space in the root block of the parent.
The expected block cost of this subproblem is the sum of the costs of the
recursive subproblems.
We evaluate the expected block cost for each partition of $K-1$
and choose the partition that minimizes the cost.

For binary trees, this dynamic program runs in $O(N B^2)$ time because,
for each of the $N (B+1)$ subproblems,
there are at most $B$ binary partitions of $K-1 < B$
and the cost to evaluate each one is $O(1)$.
However, for nonbinary trees, the time bound still has exponential dependence
on $B$ because there are $\Theta(N^K/K!)$ possible partitions of~$K$.

The final idea is to \emph{expand} each node of degree $d > 2$ and its children
into a complete binary tree on $d$ leaves (the original children).  In
addition, we assign a \emph{weight} of $0$ to all of the new (nonleaf and
nonroot) nodes in this expansion; all original nodes have weight~$1$.
A node $v$ having weight $W$ means that node $v$ occupies $W$ space
in the root block; in particular, weight $0$ means that the node is free
to include in the root block.

Expansion converts the original nonbinary tree into a weighted binary tree
having at most twice as many nodes.  Thus, the algorithm runs in $O(N B^2)$
time for any tree, as in the binary-tree case, provided that we can handle
vertex weights of $0$ and~$1$.

The dynamic program described above is easy to modify to handle vertex weights
of $0$ and~$1$.
First, when we solve a subproblem rooted at node $v$ and with root capacity
$K \geq 1$, we consider all partitions of $K-W$ instead of $K-1$,
where $W$ is the weight of node~$v$.
Second, when we solve a subproblem rooted at node $v$ and
with root capacity~$0$, we do not immediately induce a cost of the probability
of~$v$ and raise the root capacity to~$B$.  Instead, we check whether node $v$
has weight $0$, and in this case we place $v$ into the root block and
recursively solve the children subproblems with root capacity~$0$.
(If $v$ has positive weight, the algorithm is as before.)

\subsection{\boldmath Additive $1$ Error in $O(N B)$ Time}
\label{B reduction}

The following lemma allows us to reduce the time bound by a factor of $B$ by
reducing the number of branching nodes from potentially $N$ to at most $N/B$.

\begin{lemma}
  {\bf \latexcite{Dixon-Tarjan-1997, Farach-Thorup-1995-comparison}}
  Consider a rooted tree with $N$ nodes.
  If we trim the tree by removing every node whose subtree contains at
  most $B$ nodes, then the remaining tree has at most $N/B$ leaves.
\end{lemma}

\begin{proof}
  Consider a leaf in the trimmed tree.  This node was not removed because
  its subtree had strictly more than $B$ nodes.  Therefore, in the original
  tree there were at least $B$ nodes below the leaf, and all such nodes have
  since been removed.  In total, at most $N$ nodes were removed.
  Hence there are at most $N/B$ leaves in the trimmed tree,
  because we can charge each leaf to $B$ different removed nodes.
\end{proof}

Because each connected component of trimmed nodes has size at most $B$,
these nodes can be stored in a single block.  For any layout of the trimmed
tree, we can add these blocks to obtain a layout of the original tree.
If the original layout has expected block cost $C$, then the resulting layout
has an expected block cost of $C+1$ because every root-to-leaf path visits
exactly one block of trimmed nodes.  If we start with the optimal layout of the
trimmed tree, then its expected block cost $C$ is minimum and
serves as a lower bound on the expected block cost in the optimal layout of
the original tree.  Therefore, the layout of the original tree obtained from
trimming is within an additive $1$ of the optimal expected block cost.

The reduction does not reduce $N$ per se, and may not even reduce the number
of leaves in the tree, but it guarantees that the resulting tree has at most
$N/B$ leaves.  As a consequence, the tree has fewer than $N/B$ branching nodes.
This bound translates into a factor of $B$ speedup in the dynamic program
because nonbranching internal nodes can be dealt with efficiently.
Specifically, we contract every nonbranching internal node into its unique
child, adding to the weight of that child (the number of slots in a
block required by the node).  The resulting weighted tree has $N' = O(N/B)$
nodes.

Now we have a more general problem than before: each node weight is an
arbitrary nonnegative integer, possibly even larger than $B$,
not just $0$ or~$1$.
We solve this problem by
treating each subproblem with root capacity $K$ and root weight
$W$ as a subproblem with root capacity $(K-W) \bmod B$ and root weight $0$
that incurs an additional expected block cost of
$\lceil \max\{0,W-K\}/B \rceil$.
This conversion requires $O(1)$ additional time per node.
Therefore the total running time of the dynamic program is $O(N' B^2)
= O(N B)$.

In summary, we have proved the following theorem:

\begin{theorem}
  There is an $O(N B)$-time algorithm that computes a layout of a rooted tree
  with $N$ nodes and known block size $B$ whose expected block cost is within
  an additive $1$ of optimal.
\end{theorem}

Unfortunately, this reduction in problem size cannot be applied a second time.
Trimming the tree twice would not further reduce the number of leaves.
If we trim, then contract each nonbranching internal node into its
unique child, then trim again, we would reduce the number of leaves to
$O(N/B^2)$, but the additional block cost from traversing the trimmed nodes
may be much larger than~$2$ because some nodes have potentially larger weight.
Therefore we need another approach for approximation,
as offered by the next section.

\subsection{\boldmath Additive $1+\delta$ Error in $O(N)$ Time}
\label{fancy DP}

Next we show how to improve the dynamic program to run in $O(N B)$ time
by ignoring some of the subproblems, inducing an additive error of
$\delta$ for any desired $\delta > 0$.  Combined with the reduction of the
previous section, we obtain an $O(N)$-time algorithm with additive error
$1+\delta$ for any $\delta > 0$.

The new dynamic program is based on which child of each node has a larger
probability of being visited along a root-to-leaf path.
The \emph{light child} of an internal node is the child with the least
probability; the other child is the \emph{heavy child}.
(Using the above reductions to the weighted problem, we can assume that every
internal node has exactly two children.  If both children have the same
probability, we break the tie arbitrarily.)
We define the \emph{light probability} $\ell(v)$ of a node $v$ to be the
probability of its light child.
%
%
We define $x(v) = c + \max\{0, \lfloor \lg (\ell(v) \cdot N) \rfloor\}$
for a constant $c \geq 0$ to be chosen later.
Because the probability of a light child is at most half the probability of
its parent, the light probability $\ell(u)$ of a light child $u$ is at most
half the light probability $\ell(v)$ of its parent $v$,
so $x(u)$ is at least $1$ less than $x(v)$ unless $x(v) = 0$.


In the new dynamic program, we focus our attention on $1.5^{x(v)}$
\emph{important subproblems} involving the light child $u$ of a node $v$,
where the subproblems are chosen so that their solution values are
roughly evenly spaced.
To be more precise, as we increase the light child $u$'s root capacity
through all integers from $0$ to $B$, the costs of subproblem solutions
monotonically decrease.
The maximum possible variation in the cost from beginning to end
is the probability $\ell(v)$ of light child $u$, because the biggest possible
change is requiring an entire new block for root-to-leaf paths visiting~$u$.
Among the subproblem solutions for $u$, we extract a solution whose cost is
nearest and at most each of $1.5^{x(v)}$ evenly spaced values in the
length-$\ell(v)$ range of solution costs, including a value at the maximum
of the range but not at the minimum of the range.
Among multiple solutions with the same cost, we break ties by choosing the
solution that requires the smallest root capacity.
In particular, this subset of \emph{important solutions} includes the
highest-cost solution with a root capacity of~$0$.  More generally, for any
(not necessarily important) subproblem involving $u$, we can round $u$'s
root capacity down and find an important solution whose cost is at most an
additive $1/1.5^{x(v)}$ more than the solution to the original subproblem.

The new dynamic program solves the $B+1$ subproblems involving a node $v$
as follows.  First we select the $1.5^{x(v)}$ important subproblem solutions
among the $B+1$ subproblems involving the light child $u$ of $v$.  (The main
improvement in running time is that this selection can be done once for all
$B+1$ subproblems involving node~$v$.)  Then we consider separately each of
the $B+1$ subproblems involving~$v$; let $K$ denote the root capacity of $v$
in each such subproblem.  Let $W$ denote the weight of node~$v$.
For each important subproblem solution involving the light child $u$
in which the root capacity $J$ at $u$ is at most $(K-W) \bmod B$, we consider
the solution to the subproblem involving the heavy child of $v$ with the
remaining root capacity of $((K-W) \bmod B) - J$.
We take the best such combination of subproblem solutions for the light child
$u$ of $v$ and the heavy child of $v$ as our solution to this subproblem
involving~$v$.

First we claim that the running time of this algorithm is $O(N B)$.
The time to solve all subproblems involving a node $v$
is $O(B)$ for the important selection,
then $O(1.5^{x(v)})$ for each of the $B+1$ subproblems,
for a total of $O(1.5^{x(v)} B)$.
For any $x > c$, every node $v$ with $x(v)=x$ has light probability
at least $2^{x-c}/N$ (by definition of $x(v)$), and these light subtrees
cannot overlap, so their total probability is at most~$1$.
Hence the number of nodes with $x(v) = x > c$ must be at most
$1/(2^{x-c}/N) = N/2^{x-c}$.
On the other hand, the number of nodes with $x(v) = c$ is at most~$N$.
Therefore the total running time is at most
$$ O(1.5^c B) \cdot N + \sum_{x=c+1}^\infty O(1.5^x B) \cdot \frac{N}{2^x}
\ =\ O(B N). $$

Finally we claim that the total error caused by the approximation is small.
Consider the optimal solution to a subproblem involving node~$v$.
This solution makes a particular division of the root capacity of $v$
into root capacities for the children of~$v$.  If we round this partition
so that the root capacity of the light child of $v$ decreases to an important
value and the root capacity of the heavy child of $v$ increases by the same
amount, then we obtain a partition considered by the approximate dynamic
program.

Suppose that we round every optimal subproblem solution in this way.
Then we obtain an overall solution considered by the approximate dynamic
program, and the solution returned by the approximate dynamic program can be
only better than this particular solution under consideration.
Furthermore, each subproblem involving a heavy child has a larger root capacity
than the optimal, and we use the ``exact'' solution for such subproblems
(at least as far as this level is concerned).  Thus the approximate solutions
to subproblems involving heavy children can have only a smaller cost than the
optimal solution (ignoring error introduced at lower levels).
The only extra cost from rounding is the decreasing of the root capacity of
each light child.  But we already know that the additive overhead of each of
these rounding operations is at most $1/1.5^{x(v)}$ for the light child of
a node~$v$.
Any root-to-leaf path visits at most one light child of a node $v$ with
$x(v)=x$, for each $x > c$.  Also, if the leaf has probability $p \leq 1/N$,
then the path visits at most $1 + \lg (1/(N p)) = 1 + \lg (1/p) - \lg N$ light
children of nodes $v$ with $x(v)=c$.
Therefore the additive expected overhead over all root-to-leaf paths is at most
$$ \sum_{x=c+1}^\infty \frac{1}{1.5^x} \ +\ 
   \frac{1}{1.5^c} \,
     \sum_{{{\rm leaves\ with}\atop{\rm probability}}
           \atop p \leq 1/N}
       p \Big(1 + \lg (1/p) - \lg N\Big). $$
Because the sum $P$ of the $p$'s with $p \leq 1/N$ is at most $1$,
the binary entropy term $\sum_p p \lg (1/p)$ is maximized when these $p$'s are
all equal, in which case it is $P \lg N$.
This $P \lg N$ cancels with the $-P \lg N$ term,
so the expected additive overhead is in fact at most
$$ \sum_{x=c+1}^\infty \frac{1}{1.5^x} + \frac{1}{1.5^c}
   \ =\ \frac{3}{1.5^{c+1}} + \frac{1}{1.5^c}, $$
which is at most any desired $\delta > 0$ for a sufficiently large choice
of the constant~$c$.

\fi

\section{Cache-Oblivious Layout}

In this section, we develop a cache-oblivious layout whose expected block cost
is within a constant factor of the optimal layout.

\begin{theorem} \label{split-and-refine}
  The Split-and-Refine algorithm produces a cache-oblivious layout whose
  expected block cost is within a constant multiplicative factor of optimal.
  The algorithm can use any given black box that
  lays out a tree within a constant factor of the optimal expected block cost
  for a known block size $B$.
  If the running time of the black box is $T(N,B) = \Omega(N)$,
  then the Split-and-Refine running time is
  $O(\sum_{l=0}^{\lceil \lg N \rceil} T(N,2^l))$.
  \ifDP
  In particular, plugging in the approximate dynamic program from Section
  \ref{DP}, we obtain a running time of $O(N \lg N)$.
  \fi
\end{theorem}

In fact, our technique is quite general, and in addition to computing a layout
that minimizes the expected block cost, it can compute a layout that minimizes
the \emph{maximum} block cost.

\begin{theorem} \label{split-and-refine 2}
  The Split-and-Refine algorithm produces a cache-oblivious layout whose
  maximum block cost is within a constant multiplicative factor of optimal.
  The algorithm can use any given black box that
  lays out a tree within a constant factor of the optimal maximum block cost
  for a known block size $B$.
  The running times are as in Theorem \ref{split-and-refine}.
\end{theorem}

%

\subsection{Split-and-Refine Algorithm}

The basic idea of the cache-oblivious layout is to recursively combine
optimal layouts for several carefully chosen block sizes.
These layouts are computed independently,
and the block sizes are chosen so that costs of the layouts grow exponentially.
The layouts may be radically different; all we know is their order
from coarser (larger $B$) to finer (smaller $B$).
To create a recursive structure among these layouts,
we further partition each layout to be consistent with all coarser layouts.
Then we store the tree according to this recursive structure.

More precisely, our cache-oblivious layout algorithm works as follows.
For efficiency, we only consider block sizes that are powers of two.%
\footnote{%
  Restricting to powers of two speeds up the layout algorithm; if we did
  not care about speed, we could consider all possible values of~$B$.}
We begin with a block size of the smallest power of two that is at least $N$,
i.e., the \emph{hyperceiling} $\lhceil N \rhceil = 2^{\lceil \lg N \rceil}$.
Because this block size is at least $N$, the optimal partition places all nodes
in a single block, and the expected (and worst-case) search cost is~$1$.
We call this trivial partition \emph{level of detail~$0$}.
Now we repeatedly halve the current block size, and at each step we compute the
optimal partition.
We stop when we reach the coarsest partition whose expected
search cost is between $2$ and~$4$; such a partition exists by the
Smoothness Lemma (Lemma~\ref{smoothness}).
We call this partition \emph{level of detail~$1$}.
Then we continue halving the current block size, until we reach the coarsest
partition whose expected search cost is between $2$ and $4$ times the expected
search cost at level of detail~$1$.
This partition defines \emph{level of detail~$2$}.
We proceed in defining levels of detail until we reach a level of detail~$\ell$
whose block size is~$1$.
In contrast to all other levels, the expected search cost at level of
detail~$\ell$ may be less than a factor of $2$ larger than level of detail
$\ell-1$.

The levels of detail are inconsistent in the sense that a block at one level of
detail may not be wholly contained in a block at any coarser level of detail.
To define the layout recursively, we require the blocks to form a
\emph{recursive structure}: a block at one level of detail should be made up of
\emph{subblocks} at the next finer level of detail.
To achieve this property, we define the \emph{refined level of detail $i$}
to be the refinement of the partition at level of detail $i$ according to the
partitions of all coarser levels of detail $< i$.
That is, if two nodes are in different blocks at level of detail $i$, then
we separate them into different blocks at all finer levels of detail $> i$.

The recursive structure allows us to build a recursive layout as follows.
Each block at any refined level of detail is stored in a contiguous segment of
memory.  The subblocks of a block can be stored in any order as long as they
are stored contiguously.

\subsection{Running Time}

We can compute the partition at each level of detail within the claimed
running time because we call the black box precisely for block sizes $2^l$
where $l = 0, 1, \dots, \lceil \lg N \rceil$.  Each call to the black box
produces a partition on the $N$ nodes, which we represent by arbitrarily
assigning each block a unique integer between $1$ and $N$, and labeling each
node with the integer assigned to the block containing it.

{From} these unrefined partitions, it is easy to compute the
cache-oblivious layout.
To each node we assign a \emph{key} consisting of at most
$\lceil \lg N \rceil + 1$ components,
assembled from the labels from all levels of detail,
where the coarsest level of detail specifies the most significant component
of the key.
Then we sort the nodes according to their keys in $O(N \lg N)$ time
using a radix sort, and lay out the nodes in this order.
This layout automatically adheres to the refined levels of detail,
without having to compute them explicitly.

\subsection{Expected Block Cost}

We begin by analyzing the cost of the unrefined levels of detail.
Define the random variable $X_i$ to be the number of blocks along a random
root-to-leaf path in the partition defined by level of detail~$i$.
Thus, $E[X_i]$ is the expected search cost at level of detail~$i$, as above.
By construction, $E[X_{i+1}]$ is a factor between $2$ and $4$ larger
than~$E[X_i]$.

Let $B$ be the true block size of the cache, not known to the algorithm.
We focus on two levels of detail: level of detail $L-1$ whose block size is
at least~$B$, and level of detail $L$ whose block size is at most~$B$.
By the Monotonicity Lemma (Lemma~\ref{monotonicity}),
the ideal optimal partition with block size~$B$ has expected search cost
at least $E[X_{L-1}]$,
because the block size at level of detail $L-1$ is only larger than $B$.
By construction, $E[X_L]$ is at most $4$ times larger than $E[X_{L-1}]$,
and thus is at most $4$ times larger than the optimal partition with block
size~$B$.

Consider the partition defined by level of detail~$L$, which is designed for
blocks smaller than~$B$, but laid out in a memory with block size~$B$.
Each block in this partition has size at most~$B$, and hence is contained
in at most two memory blocks of size~$B$, depending on alignment.
Thus, the expected search cost measured according to block size $B$
is at most $8$ times the optimal partition with block size $B$.

\medskip
It remains to consider the additional cost introduced by refining
level of detail~$L$ by all coarser levels of detail.
Call an edge of the tree \emph{straddling at level $i$} if its endpoints lie in
different blocks at level of detail~$i$.
Connecting to the previous analysis, $X_i$ is $1$ plus the number of straddling
edges at level $i$ along a random root-to-leaf path.
The important property of straddling edges is this:
along a root-to-leaf path, the straddling edges at level~$i$ count the number
of extra memory transfers (block refinements) caused by refining level of
detail~$L$ according to coarser level of detail~$i$.
Thus, an edge spans two different refined blocks in the refined level of
detail~$L$ precisely when the edge is straddling at some level of detail
$\leq L$.

To capture these properties algebraically, define the random variable $X$ to be
the number of blocks along a random root-to-leaf path in the partition defined
by the refined level of detail~$L$.
Because $X$ counts $X_L$ (the cost of the unrefined level of detail)
as well as the extra memory transfers caused by straddling edges at levels of
detail $< L$, we have the following equation:
\begin{eqnarray*}
  X &=& (X_1 - 1) + (X_2 - 1) + \cdots + (X_{L-1} - 1) + X_L \\
    &=& X_1 + X_2 + \cdots + X_{L-1} + X_L - (L-1).
\end{eqnarray*}

Now we want to compute $E[X]$, that is, the expected search cost of the
partition defined by the refined level of detail $L$, ignoring block alignment.
By linearity of expectation,
\begin{eqnarray*}
  E[X] &=& E[X_1 + X_2 + \cdots + X_L - (L-1)] \\
       &=& E[X_1] + E[X_2] + \cdots + E[X_L] - (L-1) \\
       &\leq& E[X_1] + E[X_2] + \cdots + E[X_L].
\end{eqnarray*}
As mentioned above, each $E[X_i]$ is a factor between $2$ and $4$ more than
$E[X_{i-1}]$, so the series is geometric.
Thus,
$$E[X] \leq 2 E[X_L].$$
But we already argued that $E[X_L]$ is at most $4$ times larger than
$E[X_{L-1}]$, and that $E[X_{L-1}]$ is at most the expected search cost of the
ideal optimal partition with block size $B$.
Therefore, $E[X]$ is at most $8$ times this ideal cost,
so the refined level of detail $L$ has an expected number of memory transfers
that is at most $16$ times the optimal.
This concludes the proof of Theorem \ref{split-and-refine}.

\subsection{Minimizing the Maximum Block Cost}

Clark and Munro \cite{Clark-Munro-1996, Clark-1996}
consider an analogous tree-layout problem
in which the objective is to minimize the maximum block cost
(which is independent of any probability distribution of the leaves).
They give a simple greedy algorithm that computes the exact optimal layout
in $O(N)$ time for a known block size $B$,
The basic idea behind the algorithm is to build a partition bottom-up,
starting with the leaves in their own blocks, and merging blocks locally
as much as possible while giving priority to the most expensive subtrees
(those with highest maximum block cost).

We can use this layout algorithm for known $B$ as an alternative subroutine in
the cache-oblivious Split-and-Refine algorithm.  The algorithm behaves
correctly as before because of an analogous Smoothness Lemma for the metric of
maximum block cost (exactly the same proof of Lemma \ref{smoothness} applies).
The analysis also proceeds as before, except that we replace $X_i$ and
$E[X_i]$ by $Y_i$, which is defined as the maximum number of blocks along any
root-to-leaf path in the block partition defined by level of detail~$i$.
There is no longer any randomization or expectation,
but otherwise the proof is identical.

\section{Conclusion}

In this paper, we developed cache-oblivious layouts of fixed-topology trees,
whose performance is within a constant factor of the optimal layout
with known block size.
The running time of the layout algorithm is dominated by $O(\lg N)$ calls to
any given layout algorithm for known block size.
\ifDP
In particular, applying our efficient dynamic-programming algorithms from
Section \ref{DP}, we obtain a running time of $O(N \lg N)$.
\fi
We also showed that two natural greedy strategies have a performance
ratio between $\Omega(\log B / \log \log B)$ and $O(\log B)$
when compared to the optimal layout.

The main open problems are to what extent fixed-topology tree layouts can
be made dynamic and/or self-adjusting, both with known block size and
in the cache-oblivious setting.  By \emph{dynamic} we mean the ability to
add and delete leaves, and to redistribute probability from leaf to leaf.
A \emph{self-adjusting} data structure would adapt (as in splay trees)
to an online search distribution without a priori knowledge of the
distribution.

\section*{Acknowledgments}

We thank Martin Demaine, Stefan Langerman, and Ian Munro for many helpful
discussions.

\let\realbibitem=\bibitem
\def\bibitem{\par \vspace{-1.2ex}\realbibitem}

\small
\bibliography{algs,biology,cache,coding,parallel,searchtrees,strings,succinct,new}

\newcommand{\etalchar}[1]{$^{#1}$}
\begin{thebibliography}{BCDFC02}

\bibitem[ABD{\etalchar{+}}02]{Arge-Bender-Demaine-Holland-Minkley-Munro-2002}
Lars Arge, Michael~A. Bender, Erik~D. Demaine, Bryan Holland-Minkley, and
  J.~Ian Munro.
\newblock Cache-oblivious priority queue and graph algorithm applications.
\newblock In {\em Proceedings of the 34th Annual ACM Symposium on Theory of
  Computing}, pages 268--276, Montr\'eal, Canada, May 2002.

\bibitem[AHU74]{Aho-Hopcroft-Ullman-1974}
Alfred~V. Aho, John~E. Hopcroft, and Jeffrey~D. Ullman.
\newblock {\em The Design and Analysis of Computer Algorithms}.
\newblock Addison-Wesley, 1974.

\bibitem[BCDFC02]{Bender-Cole-Demaine-Farach-Colton-2002-scans}
Michael~A. Bender, Richard Cole, Erik~D. Demaine, and Martin Farach-Colton.
\newblock Scanning and traversing: Maintaining data for traversals in a memory
  hierarchy.
\newblock In {\em Proceedings of the 10th Annual European Symposium on
  Algorithms}, volume 2461 of {\em Lecture Notes in Computer Science}, pages
  139--151, Rome, Italy, September 2002.

\bibitem[BCR02]{Bender-Cole-Raman-2002}
Michael~A. Bender, Richard Cole, and Rajeev Raman.
\newblock Exponential structures for efficient cache-oblivious algorithms.
\newblock In {\em Proceedings of the 29th International Colloquium on Automata,
  Languages and Programming}, volume 2380 of {\em Lecture Notes in Computer
  Science}, pages 195--207, M\'alaga, Spain, July 2002.

\bibitem[BDFC00]{Bender-Demaine-Farach-Colton-2000}
Michael~A. Bender, Erik~D. Demaine, and Martin Farach-Colton.
\newblock Cache-oblivious {B}-trees.
\newblock In {\em Proceedings of the 41st Annual Symposium on Foundations of
  Computer Science}, pages 399--409, Redondo Beach, California, November 2000.

\bibitem[BDFC02]{Bender-Demaine-Farach-Colton-2002-tree-layout}
Michael~A. Bender, Erik~D. Demaine, and Martin Farach-Colton.
\newblock Efficient tree layout in a multilevel memory hierarchy.
\newblock In {\em Proceedings of the 10th Annual European Symposium on
  Algorithms}, volume 2461 of {\em Lecture Notes in Computer Science}, pages
  165--173, Rome, Italy, September 2002.

\bibitem[BDIW02]{Bender-Duan-Iacono-Wu-2002}
Michael~A. Bender, Ziyang Duan, John Iacono, and Jing Wu.
\newblock A locality-preserving cache-oblivious dynamic dictionary.
\newblock In {\em Proceedings of the 13th Annual ACM-SIAM Symposium on Discrete
  Algorithms}, pages 29--38, San Francisco, California, January 2002.

\bibitem[BFJ02]{Brodal-Fagerberg-Jacob-2002}
Gerth~St{\o}lting Brodal, Rolf Fagerberg, and Riko Jacob.
\newblock Cache oblivious search trees via binary trees of small height.
\newblock In {\em Proceedings of the 13th Annual ACM-SIAM Symposium on Discrete
  Algorithms}, pages 39--48, San Francisco, California, January 2002.

\bibitem[BM72]{Bayer-McCreight-1972}
Rudolf Bayer and Edward~M. McCreight.
\newblock Organization and maintenance of large ordered indexes.
\newblock {\em Acta Informatica}, 1(3):173--189, February 1972.

\bibitem[Cla96]{Clark-1996}
David Clark.
\newblock {\em Compact Pat Trees}.
\newblock PhD thesis, University of Waterloo, 1996.

\bibitem[CM96]{Clark-Munro-1996}
David~R. Clark and J.~Ian Munro.
\newblock Efficient suffix trees on secondary storage.
\newblock In {\em Proceedings of the 7th Annual ACM-SIAM Symposium on Discrete
  Algorithms}, pages 383--391, Atlanta, January 1996.

\bibitem[DT97]{Dixon-Tarjan-1997}
Brandon Dixon and Robert~Endre Tarjan.
\newblock Optimal parallel verification of minimum spanning trees in
  logarithmic time.
\newblock {\em Algorithmica}, 17(1):11--18, 1997.

\bibitem[FLPR99]{Frigo-Leiserson-Prokop-Ramachandran-1999}
Matteo Frigo, Charles~E. Leiserson, Harald Prokop, and Sridhar Ramachandran.
\newblock Cache-oblivious algorithms.
\newblock In {\em Proceedings of the 40th Annual Symposium on Foundations of
  Computer Science}, pages 285--297, New York, October 1999.

\bibitem[FT95]{Farach-Thorup-1995-comparison}
Martin Farach and Mikkel Thorup.
\newblock Fast comparison of evolutionary trees.
\newblock {\em Information and Computation}, 123(1):29--37, 1995.

\bibitem[GI95]{Gil-Itai-1995}
Joseph Gil and Alon Itai.
\newblock Packing trees.
\newblock In {\em Proceedings of the 3rd Annual European Symposium on
  Algorithms ({ESA})}, pages 113--127, 1995.

\bibitem[GI99]{Gil-Itai-1999}
Joseph Gil and Alon Itai.
\newblock How to pack trees.
\newblock {\em Journal of Algorithms}, 32(2):108--132, 1999.

\bibitem[HT71]{Hu-Tucker-1971-SIAM}
T.~C. Hu and A.~C. Tucker.
\newblock Optimal computer search trees and variable-length alphabetic codes.
\newblock {\em SIAM Journal on Applied Mathematics}, 21(4):514--532, December
  1971.

\bibitem[Huf52]{Huffman-1952}
David~A. Huffman.
\newblock A method for the construction of minimum-redundancy codes.
\newblock {\em Proceedings of the IRE}, 40(9):1098--1101, 1952.

\bibitem[Iac01]{Iacono-2001-splay}
John Iacono.
\newblock Alternatives to splay trees with {$O(\log n)$} worst-case access
  times.
\newblock In {\em Proceedings of the 12th Annual ACM-SIAM Symposium on Discrete
  Algorithms}, pages 516--522, Washington, D.C., January 2001.

\bibitem[Knu68]{Knuth-1968-volume-3}
Donald~E. Knuth.
\newblock {\em The Art of Computer Programming}, volume 3 (Sorting and
  Searching).
\newblock Addison-Wesley, 1968.

\bibitem[Pro99]{Prokop-1999}
Harald Prokop.
\newblock Cache-oblivious algorithms.
\newblock Master's thesis, Massachusetts Institute of Technology, Cambridge,
  MA, June 1999.

\bibitem[RCR01]{Rahman-Cole-Raman-2001}
Naila Rahman, Richard Cole, and Rajeev Raman.
\newblock Optimised predecessor data structures for internal memory.
\newblock In {\em Proceedings of the 5th International Workshop on Algorithm
  Engineering}, volume 2141 of {\em Lecture Notes in Computer Science}, pages
  67--78, Aarhus, Denmark, August 2001.

\bibitem[ST85]{Sleator-Tarjan-1985-splay}
Daniel~Dominic Sleator and Robert~Endre Tarjan.
\newblock Self-adjusting binary search trees.
\newblock {\em Journal of the ACM}, 32(3):652--686, July 1985.

\bibitem[Thi01]{Thite-2001}
Shripad Thite.
\newblock Optimum binary search trees on the hierarchical memory model.
\newblock Master's thesis, Department of Computer Science, University of
  Illinios at Urbana-Champaign, 2001.

\end{thebibliography}
\bibliographystyle{alpha}

\end{document}